\documentclass[
 reprint,
 amsmath,amssymb,
 aps,
prb,
]{revtex4-2}

\usepackage{array}
\usepackage{makecell}
\usepackage{setspace}
\usepackage{graphicx}
\usepackage{dcolumn}
\usepackage{bm}
\usepackage{url}
\usepackage[colorlinks=true,linkcolor=blue,citecolor=blue,urlcolor=blue]{hyperref}

\begin{document}

\preprint{APS/123-QED}

\title{Phonon-assisted optical absorption of SiC polytypes from first principles}

\author{Xiao Zhang}
\author{Emmanouil Kioupakis}%
\email{kioup@umich.edu}
\affiliation{%
 The University of Michigan, Ann Arbor, Department of Materials Science and Engineering, Ann Arbor, 48109, USA
}%

\begin{abstract}
Silicon carbide (SiC) is an indirect-gap semiconductor material widely used in electronic and optoelectronic applications. While experimental measurements of the phonon-assisted absorption coefficient of SiC across its indirect gap have existed for more than fifty years, theoretical investigations of phonon-assisted absorption have been hampered by their excessive computational cost. In this work, we calculate the phonon-assisted temperature-dependent optical absorption spectra of the commonly occurring SiC polytypes (3C, 2H, 4H and 6H), using first-principles approaches based on density functional theory and related techniques. We show that our results agree with experimentally determined absorption coefficients in the spectral region between the direct and indirect band gaps.  The temperature dependence of the spectra can be well-predicted with taking the temperature-dependence of the band gaps into account. Lastly, we compare the spectra obtained with second-order perturbation theory to those determined by the special displacement method, and we show that the full consideration of the electronic energy renormalization due to temperature is important to further improve the prediction of the phonon-assisted absorption in SiC. Our insights can be applied to predict the optical spectra of the less common SiC polytypes and other indirect-gap semiconductors in general.
\end{abstract}

\maketitle

\section{\label{sec:intro}Introduction}
Silicon carbide (SiC) is an indirect gap semiconductor material that has been well established and widely used in many electronic and optoelectronic devices. It has a variety of structural polytypes, including cubic 3C; hexagonal 2H, 4H, 6H and several Rhombohedral structures (9R, 15R, etc.) \cite{chien1994terrace,feng2013sic,Yaghoubi2018}, that enable a broad range of applications. Its 6H polytype is one of the first to be used to create blue light emitting diodes (LEDs) that enabled full color LEDs \cite{edmond19976h}, though the indirect band gap limits the efficiency. Absorbing short-wavelength light much more efficiently, the material is also used as UV-sensitive diodes in flame sensors\cite{edmond19976h,BROWN1998755,przybylko1993developments}. More recently, SiC is widely used as substrate material for applications such as extreme condition transistors \cite{AIDA2012S41,Okumura_2006,kimoto2003high}, tunable photo-detecting devices \cite{gao2019}, and advanced UV-detectors\cite{liu2019transferable,li2019highly,huang2020ultraviolet}. Recent developments in crystal growth also allowed novel structures such as thin films, nanoparticles, heterostructures, etc., for more advanced applications\cite{kim2017recent,mousa2019performance,ahmed2020determination,bradford2020synthesis}. Due to its early discovery as well as the wide applications and interest, it is now one of the most well studied semiconductor materials from both an experimental and a theoretical point of view. 

Understanding the similarities and differences between the structural polytypes is crucial for practical applications. The most common SiC polytypes for application purposes are the cubic structure (3C) and two hexagonal structures with different stackings (4H and 6H). It has been shown both by experimental and theoretical studies \cite{scalise2019temperature} that at low temperatures, SiC tends to form the cubic 3C structure while at higher temperatures, 4H and 6H become the more energetically favorable structures. In addition to 4H and 6H, the 2H structure of SiC has also been observed, although rarely due to being energetically less favorable. Due to its rare occurrence, the optoelectronic properties of the 2H polytype have not been studied thoroughly from either a theoretical or an experimental perspective, and previous studies are typically limited to only structural properties and basic electronic properties such as the band gap. The rhombohedral structure of SiC is less commonly seen but can also be grown on various temperatures and substrates \cite{schomer1999significantly,mourya2018structural,yano2000mosfet}, and has been shown to demonstrate great potential in MOSFET applications \cite{schomer1999significantly,yano2000mosfet}. Overall, the natural occurrence of these different polytypes make it an important task to study them consistently from a first-principles perspective consistently to understand the properties of the material better, and potentially discover new opportunities for applications.

Despite being well studied for many polytypes, the indirect optical properties of SiC, which are important considering its applications in optoelectronic devices, have never been thoroughly studied with first principles techniques. Experimental characterization of its indirect gap and optical absorption in the spectral region between the indirect and direct band gaps already exists ever since the 1960s \cite{choyke1969optical,o1960silicon}. There has also been a considerable number of theoretical calculations of optical properties for both bulk SiC and nanostructures, however, always focusing on direct transitions \cite{zhang2017first,majidi2017first,xie2003first,lee1994first,niu2019first,song2009electronic,xu2016controlling}. The commonly used theoretical spectroscopy approaches based on the density functional theory lack descriptions of electron-phonon interactions. As a result, momentum transfer is neglected in considering optical absorption and such approaches are only able to consider direct transitions. Yet indirect optical absorption plays an important role in the application of SiC as the difference between the direct gap and the indirect gap can range from about 1 eV up to more than 3 eV, depending on the polytype\cite{feng2013sic}. As an example, although SiC was one of the first LED materials for blue light mission, it can only emit visible light via phonon-assisted transitions across its indirect band gap. Although being considerably weaker compared to direct transitions, the phonon-assisted contribution enabled its application in optoelectronic devices. Therefore, the lack of theoretical studies on the indirect optical properties of SiC significantly limits our ability to understand the material and its further potential applications in optoelectronic devices.

There are numerous computational challenges on evaluating phonon-assisted optical properties from a theoretical perspective, and tools to model them have only emerged in recent years. One significantly challenge is the necessity of dense sampling of the electronic states in the first Brillouin zone, due to the broad energy range of photon frequencies, combined with the large number of processes needed to be considered in second-order transition. While the latter makes the problem more complicated by nature, the task of studying electron-phonon properties with dense electronic and phonon Brillouin-zone sampling grids can be overcome by the approach of maximally localized Wannier function interpolation.\cite{RevModPhys.84.1419} The Wannier interpolation starts from a coarse grid of electronic states, finds a set of maximally localized Wannier functions to represent the electronic wave functions in the real space, and utilize such set of Wannier functions to construct the electronic wave functions in a dense grid of the electronic states. It has been shown that Wannier interpolation can result in accurate interpolations for both the electronic structures and electron-phonon coupling matrix elements in a broad range of materials\cite{giustino2017electron}. It thus becomes feasible to calculate electron-phonon properties on coarse grid with reasonable computational cost, then utilize Wannier interpolation to determine the phonon-assisted optical absorption spectra with adequate spectral resolution.

In this work, we perform first-principles calculations to understand the phonon-assisted optical absorption spectra for common SiC polytypes (3C, 2H, 4H, 6H, and 15R). As in experimental measurements, the absorption coefficients are mostly measured along the c-axis ($E\perp c$), thus in our work we report and compare the calculated absorption coefficient for the ordinary direction as well. By combining standard first-principles approaches (DFT, Many-body perturbation theory) for electronic structure and direct optical properties along with Wannier interpolation for electron-phonon coupling, we show that our calculated indirect optical spectra are in good agreement with available experimental measurements. We further predict the phonon-assisted spectra for polytypes for which experimental data are lacking. Our approach well predicts the temperature dependence of the indirect optical spectra, which can benefit the applications of SiC in extreme conditions. Further, by comparing to special displacement method (SDM) \cite{zacharias2016one,zacharias2015stochastic,zacharias2020theory}, we show that to further improve the predictions of indirect optical properties s a function of temperature, both the effects of temperatures on the optical spectra due to changes in phonon occupation factors   and the effects of temperatures on the electronic energy renormalization are important.

\section{\label{sec:method}Computational approach}

To calculate the structural, electronic, and optical properties of the SiC polytypes, we performed first-principles calculations based on density functional theory. The calculation is carried out with the Quantum Espresso\cite{giannozzi2009quantum,giannozzi2017advanced} package using the Perdew-Burke-Ernzerhof (PBE)\cite{PhysRevLett.77.3865} approximation for the exchange-correlation functional and the SG15 Optimized Norm-Conserving Vanderbilt (ONCV) pseudopotentials.\cite{Hamann13optimized,SCHLIPF201536} The wave functions are expanded into plane waves up to an energy cutoff of 60 Ry. Structural relaxation is done for all polytypes until all the components of the forces on the atoms are smaller than 10$^{-5}$ Ry/Bohr. Converged ground-state calculations for each polytype were performed with Brillouin-zone (BZ) sampling grids of $8\times8\times8$ for 3C, $12\times12\times8$ for 2H, $12\times12\times4$ for 4H, $10\times10\times2$ for 6H, and $8\times8\times8$ for 15R. In all cases, the grid is shifted by half a grid spacing to improve convergence. 

To correct the underestimation of the band gap by PBE, we calculated quasiparticle energies with the one-shot $GW$ method ($G_0 W_0$) using the BerkeleyGW package.\cite{hybertsen86,DESLIPPE20121269} The static dielectric matrix is evaluated in the random phase approximation and extended to finite frequency with the generalized plasmon-pole model by Hybertsen and Louie.\cite{hybertsen86} The static remainder approach is used to reduce the number of empty orbitals required.\cite{PhysRevB.87.165124} The $GW$ calculations are performed with BZ-sampling points grid of $6\times6\times6$ for 3C, $6\times6\times4$ for 2H, $6\times6\times3$ for 4H, $6\times6\times2$ and $6\times6\times6$ for 15R. For all polytypes, the quasiparticle energies were interpolated with the maximally localized Wannier function method\cite{RevModPhys.84.1419} to obtain accurate quasiparticle band structures as well as velocity and electron-phonon coupling matrix elements for fine BZ sampling grids in subsequent optical property calculations.

Phonon-related properties are calculated with density functional perturbation theory\cite{RevModPhys.73.515} as implemented in Quantum Espresso and interpolated to fine BZ-sampling grids with the maximally localized Wannier function method as implemented in the Electron-Phonon Wannier (EPW) package. \cite{Giustino2007,Ponce2016,verdi15,Noffsinger2012} The coarse phonon BZ-sampling grids used for the different polytypes are $6\times6\times6$ for 3C, $6\times6\times4$ for 2H, $6\times6\times3$ for 4H, $6\times6\times2$ for 6H, and $6\times6\times6$ for 15R. We note that the choices of the coarse BZ samplings are made to ensure both accurate Wannier interpolations as well as retain reasonable computational cost. Since SiC is a polar material, the F\"{o}hlich interaction is considered analytically via a long-range term when calculating the electron-phonon matrix elements.\cite{verdi15}
The phonon-assisted optical absorption spectra are determined with second-order time-dependent perturbation theory. Phonon-assisted optical absorption processes are second-order processes in which electrons are not only excited vertically in a band-structure diagram through energy transfer by photon absorption, but also horizontally through momentum transfer with the emission or absorption of phonons. The imaginary part of the dielectric function is derived from second-order time-dependent perturbation theory as\cite{Noffsinger2012,PhysRevB.106.205203,giustino2017electron}:
\begin{equation}
\label{eqn:eps2}
\begin{aligned}
    \varepsilon_2(\omega)=&\frac{8\pi^2 e^2}{\Omega\omega^2}\frac{1}{N_{\mathbf{k}}N_{\mathbf{q}}}\sum_{\nu ij\mathbf{k}\mathbf{q}}|\mathbf{e}\cdot[\mathbf{S}_{1,ij\nu}(\mathbf{k,q})+\mathbf{S}_{2,ij\nu}(\mathbf{k,q})]|^2 \\ &\times P_{ij}(\mathbf{k,q})\delta(\epsilon_{j,\mathbf{k+q}}-\epsilon_{i,\mathbf{k}}-\hbar\omega\pm\hbar\omega_{\nu\mathbf{q}}),
\end{aligned}
\end{equation}
where the upper (lower) sign represents the phonon emission (absorption) process. $\Omega$ is the volume of the unit cell. $\mathbf{k}$ and $\mathbf{q}$ are the electron and phonon wave-vectors. $N_{\mathbf{k}}$ and $N_{\mathbf{q}}$ are the number of $\mathbf{k}$ and $\mathbf{q}$ points in the Brillouin zone sampling. $i,j$ label the band indices and $\nu$ labels phonon modes. $\epsilon_{j,\mathbf{k+q}}$ and $\epsilon_{i,\mathbf{k}}$ are the electronic energies of state $(j,\mathbf{k+q})$ and $(i,\mathbf{k})$, respectively. The generalized matrix elements for the two possible scattering process is described by the terms $\mathbf{S}_{1,ij\nu}(\mathbf{k,q})$ (i.e., photon absorption from electronic state $(i,\mathbf{k})$ to intermediate state $(m,\mathbf{k})$ followed by electron-phonon scattering from $(m,\mathbf{k})$ to final state $(j,\mathbf{k+q})$ and $\mathbf{S}_{2,ij\nu}(\mathbf{k,q})$ (i.e., electron-phonon scattering from $(i,\mathbf{k})$ to $(m,\mathbf{k+q})$ followed by photon absorption from  $(m,\mathbf{k+q})$ to final state $(j,\mathbf{k+q})$. Mathematically, the two terms are given by:\cite{Noffsinger2012,PhysRevB.106.205203}
\begin{equation}\label{eq:s1}
   \mathbf{S}_{1,ij\nu}(\mathbf{k,\mathbf{q}})=\sum_m \frac{\textbf{v}_{im}(\mathbf{k})g_{mj,\nu}(\mathbf{k,q})}{\epsilon_{m,\mathbf{k}}-\epsilon_{i,\mathbf{k}}-\hbar\omega+i\eta},
\end{equation}
and
\begin{equation}\label{eq:s2}
   \mathbf{S}_{2,ij\nu}(\mathbf{k,\mathbf{q}})=\sum_m \frac{g_{im,\nu}(\mathbf{k,q})\textbf{v}_{mj}(\mathbf{k+q})}{\epsilon_{m,\mathbf{k+q}}-\epsilon_{i,\mathbf{k}}\pm\hbar\omega_{\nu\mathbf{q}}+i\eta}.
\end{equation}
In the equations above, $v_{ij}(\mathbf{k})$ represents the velocity matrix element for the optical transition between bands $i$ and $j$ at $\mathbf{k}$, and $g_{ij,\nu}(\mathbf{k,q})$ represents the electron-phonon matrix element describing the scattering process from electronic state $(i,\mathbf{k})$ to electronic state $(j,\mathbf{k+q})$ through phonon mode $\nu\mathbf{q}$. In addition, the occupation factor $P_{ij}(\mathbf{k,q})$ in Eq.\eqref{eqn:eps2} is given by combining the Bose-Einstein occupation factor of the phonons $n_{\nu\mathbf{q}}$ and the Fermi occupation factor of the electrons $f_{n\mathbf{k}}$. Considering energy conservation, we obtain the occupation factor for the phonon absorption $P_{a,ij}(\mathbf{k,q})$ and phonon emission $P_{e,ij}(\mathbf{k,q})$ process as:
\begin{equation}
\label{eqn:pa}
\begin{aligned}
    P_{a,ij}(\mathbf{k,q})=&n_{\mathbf{\nu\mathbf{q}}}\times f_{i,\mathbf{k}}\times(1-f_{j,\mathbf{k+q}})\\&-(n_{\nu\mathbf{q}}+1)\times(1-f_{i,\mathbf{k}})\times f_{j,\mathbf{k+q}},
\end{aligned}
\end{equation}
and
\begin{equation}
\label{eqn:pe}
\begin{aligned}
    P_{e,ij}(\mathbf{k,q})=&(n_{\mathbf{\nu\mathbf{q}}}+1)\times f_{i,\mathbf{k}}\times(1-f_{j,\mathbf{k+q}})\\&-n_{\nu\mathbf{q}}\times(1-f_{i,\mathbf{k}})\times f_{j,\mathbf{k+q}}.
\end{aligned}
\end{equation}
In addition, $\eta$ in Eq.\eqref{eq:s1} and Eq.\eqref{eq:s2}  is a numerical broadening parameter to prevent singularities induced by zeros in the denominator that occurs when direct transitions are allowed. In our work, we focus on the optical spectra between the indirect band gap and the direct band gap that are not affected by $\eta$.  

Eq.\eqref{eqn:eps2} to \eqref{eqn:pe} serve as the foundation of calculating phonon-assisted optical properties from first principles utilizing second-order perturbation theory. In this work, we calculated $\varepsilon_2(\omega)$ resulting from phonon-assisted optical absorption using the maximally localized Wannier function method to interpolate the quasiparticle energies, velocity matrix elements, phonon frequencies, and electron-phonon-coupling matrix elements onto fine BZ-sampling grids needed to converge the spectra. The fine electronic k-point and phonon q-point sampling grids we employed are $32\times32\times32$ for 3C, $24\times24\times16$ for 2H, $24\times24\times12$ for 4H, $24\times24\times8$ for 6H and $24\times24\times24$ for 15R. The delta function in Eq.\eqref{eqn:eps2} is approximated by a Gaussian function with a broadening of 0.05 eV to resolve fine features close to the absorption edge. 

The direct part of the optical spectra is calculated including electron-hole Coulomb interactions by solving the Bethe-Salpeter equation for the optical polarization function, implemented in the BerkeleyGW package.\cite{rohlfing00,hybertsen86,DESLIPPE20121269} The electron-hole interaction kernel is calculated on homogeneous electronic k-grid as in the $GW$ calculations, and interpolated to the following finer grid through considering the wavefunction overlap between the fine grids and coarse grids.\cite{DESLIPPE20121269} A small arbitrary shift is applied to all of the fine grids to ensure smooth spectra: $12\times12\times12$ for 3C, $8\times8\times6$ for 2H, $8\times8\times3$ for 4H, $9\times9\times3$ for 6H and $8\times8\times8$ for 15R. For the direct part of the spectra, a Gaussian function with a broadening of 0.15 eV is used to model the delta function. The imaginary part of the dielectric function is calculated for a total number of bands of two times the number of valence bands for each polytype (corresponds to a maximum energy of at least 30 eV and $\sim$15 eV above valence band maximum), and the real part of the dielectric function is evaluated by summing the direct and phonon-assisted part of $\varepsilon_2(\omega)$ and utilizing the Kramers-Kronig relationship. The combined real and imaginary parts of the dielectric function are then used to evaluate the refractive indices ($n_r$) and absorption coefficients (See appendix section \ref{secapp:nr} for the comparison between the calculated $n_r$ and experimental measurements). We mention that the imaginary part of the dielectric function resulting from the phonon-assisted process is typically a few orders of magnitude smaller than that resulting from direct transitions. As a result, the contribution of the phonon-assisted process to the overall integral in the Kramers-Kronig relationship is usually negligible, i.e., the difference between the refractive index obtained with and without considering the phonon-assisted contribution is small.

\begin{table*}[!ht]
    \centering
    \caption{Lattice constants (in \r{A}) of the investigated SiC polytypes as calculated in the present study and compared to previous theoretical and experimental studies. For the 3C structure, the lattice parameters are also converted to the equivalent 3H structure and shown in parentheses. Our calculated values are in excellent agreement with experimental data for all polytypes with a maximum difference of 0.8\%. 
    }
    \setlength{\tabcolsep}{4.5pt}
    \begin{tabular}{cccccccc} \hline
    & & \multicolumn{2}{c}{This work} &\multicolumn{2}{c}{Previous theory} & \multicolumn{2}{c}{Experiment} \\ 
    Polymorph & Space group& $a$  &$c$& $a$ & $c$ & $a$ &$c$ \\ \hline
        3C (3H) & F$\bar{4}$3m (206) & 4.382 (3.099) & - &4.372 (3.091)\cite{alkhaldi2019crystal}&-&4.358 (3.082)\cite{feng2013sic}&-\\
        2H & P6$_3$mc (186) & 3.094 & 5.077 &3.086\cite{alkhaldi2019crystal}&5.065\cite{alkhaldi2019crystal}&3.076\cite{feng2013sic}&5.048\cite{feng2013sic}\\
        4H & P6$_3$mc (186) & 3.096 & 10.135 &3.094\cite{alkhaldi2019crystal}&10.129\cite{alkhaldi2019crystal}&3.073\cite{feng2013sic}&10.053\cite{feng2013sic} \\
        6H & P6$_3$mc (186) & 3.097 &15.194 &3.094\cite{alkhaldi2019crystal}&15.185\cite{alkhaldi2019crystal}&3.081\cite{feng2013sic}&15.117\cite{feng2013sic} \\
        15R& R3m (160) &3.090&37.910 &3.082\cite{ching2006electronic}&37.796\cite{ching2006electronic}&3.07-3.08\cite{jarrendahl1998materials}&37.30-37.80\cite{jarrendahl1998materials}\\ \hline
    \end{tabular}
    \label{tab:str}
\end{table*}

In addition to the approach derived above from second-order perturbation theory, phonon-assisted optical properties can be obtained by properly displacing the atoms in supercells according to the eigenvalues and eigenmodes of the dynamical matrix and calculating direct transitions in the resulting distorted supercells, namely the special displacement method (SDM).\cite{zacharias2016one,zacharias2015stochastic,zacharias2020theory} The benefit of the SDM approach is that the phonon-induced band gap renormalization, as well as the temperature dependence of the band structure, are taken into consideration by nature with the construction of the specific supercell. The SDM approach uses the vibrational eigenmodes and eigenfrequencies obtained from density functional perturbation theory to generate the optimal supercell.\cite{zacharias2016one} In the SDM, phonon-related effects are directly captured via the construction of the supercell using the specific atomic displacements, the indirect part of the optical spectra can be directly calculated out using standard approaches utilizing first-order Fermi's Golden rule. In our work, we adopted the SDM approach to construct $3\times3\times3$, $4\times4\times4$ and $6\times6\times6$ supercells for the 3C polytype to compare the SDM approach with the standard approach. We note that in the SDM approach, due to the requirement of calculations for large supercells, evaluating electron-hole interactions via the BSE formalism becomes computationally expensive, thus our optical calculation is limited to the independent-particle picture.

\section{\label{sec:result}Result and discussion}
\subsection{\label{sec:groundstate}Structural, electronic and phonon properties}

The calculated structural parameters for all polytypes considered in this study are listed in Table \ref{tab:str}, and are in good agreement with previously reported computational results and experimental measurements. Our results for the lattice constants of all polytypes agree well with literature values using the same type of exchange-correlation functional  (PBE), with the maximum difference not exceeding 0.3\%.
For better comparison to the hexagonal polytypes, we convert the cubic lattice parameter of the 3C structure to the equivalent hexagonal 3H structure\cite{feng2013sic}, i.e., $a_{3H}=\frac{1}{\sqrt{2}}a_{3C}$, which is shown in the parentheses in \ref{tab:str}. For hexagonal structures, our calculations show a minor expansion of the a lattice constant and $a$ contraction of the $c$ lattice constant as the number of hexagonal layers increases. Their variation, however, is on the order of the variation of the experimental lattice constant reported by different authors.\cite{feng2013sic,park_structural_1994,stockmeier_lattice_2009,madelung_semiconductors_2012} Overall, the theoretical lattice constants are consistently overestimated in comparison to experiment, which is expected for GGA-type of exchange-correlation functionals due to its known underbinding of chemical bonds. However, the calculated trend of the variation of lattice parameters across different polytypes agrees well with experimental measurements. The maximum overestimation of the calculated lattice constant compares to the experimental value is only 0.8\%.
\begin{figure*}[!ht]
\centering
\includegraphics[width=\textwidth]{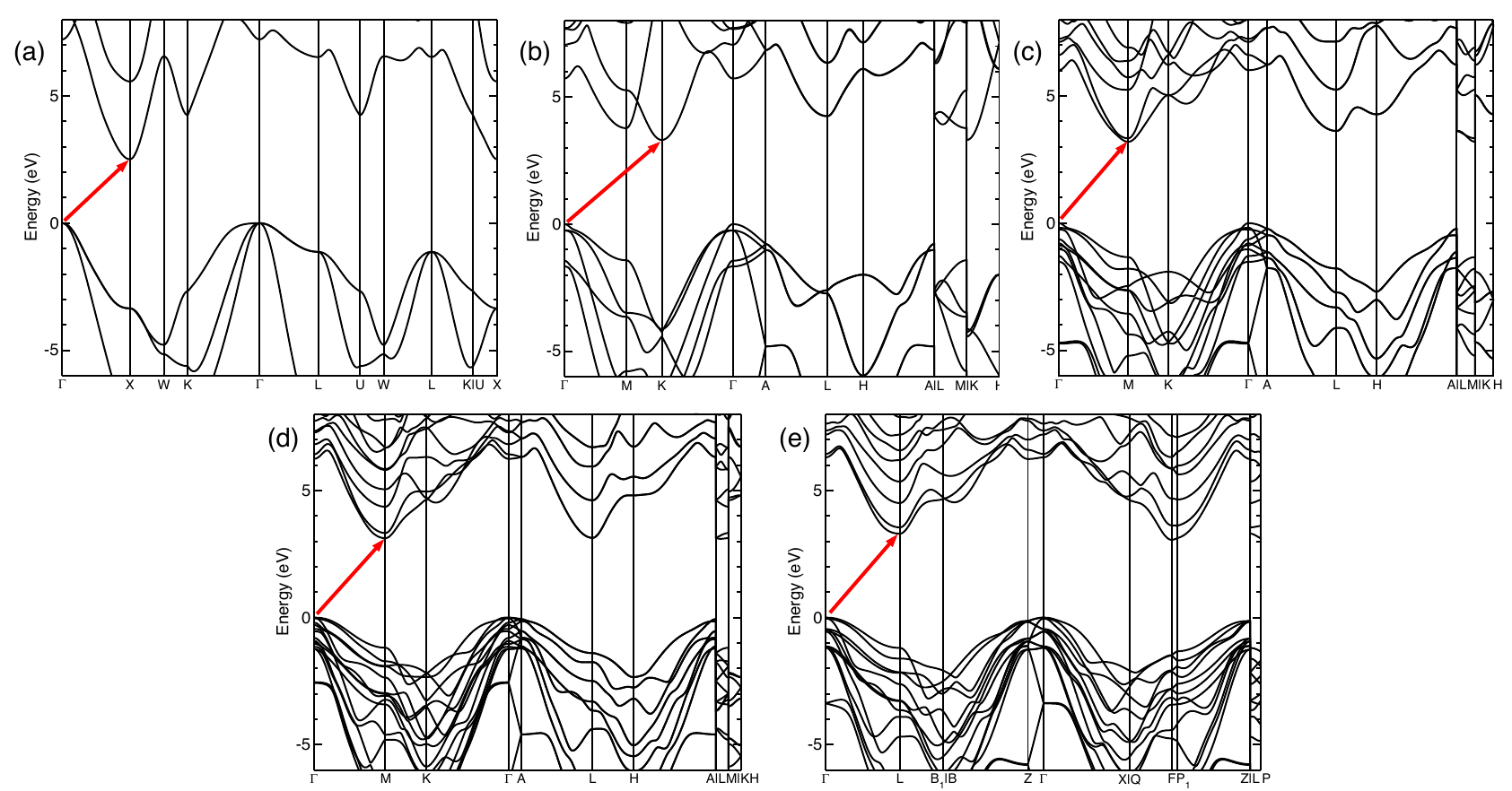}
\caption{\label{fig:bandstr} Quasiparticle band structures of the five investigated SiC polytypes: (a) 3C, (b) 2H, (c) 4H, (d) 6H and (e) 15R calculated with the $GW$ method and interpolated with the Maximally Localized Wannier Function method. All polytypes are indirect-gap semiconductors (the indirect gap is marked with red arrows).}
\end{figure*}
The electronic band structures of all investigated polytypes of SiC are shown in Figure \ref{fig:bandstr}. From the figure it can be clearly seen that all polytypes exhibit indirect band gaps that are much smaller than their direct band gaps. Overall, we see that the band structure of the 3C structures shows clearly the smallest indirect band gap and the largest direct gap compared to other polytypes. It is interesting to notice that the 2H structure exhibits a different feature that the other hexagonal structures (4H and 6H): whereas the conduction band minimum (CBM) of 4H and 6H SiC occurs at the M point of the BZ, the CBM of 2H SiC occurs at the K point, while the local minimum at M lies about 0.4 eV higher in energy. Later, we show that this difference between the 2H structure and the other hexagonal polytypes induce a unique double bump-like feature in its indirect optical absorption as described in section \ref{sec:indabs}.  

\begin{table*}[!ht]
\vspace{-0.4cm}
    \centering
    \caption{Calculated band gaps (in eV) of the five investigated SiC polytypes both within PBE and with the $GW$ approximation, in comparison to previous theoretical and experimental results. Our data are in good agreement with previous work with a maximum difference of 0.15 eV.
    }
    \setlength{\tabcolsep}{5pt}
    \begin{tabular}{cccccc} \hline
    Polytype & This work, PBE  &  Previous theory, GGA & This work, $GW$& Previous theory, $GW$ & Experiment\\ \hline 
    3C  &   1.373   &   1.391\cite{alkhaldi2019crystal},1.410\cite{huang2015thermoelectric}&   2.514     &   2.38\cite{brudnyi2012electronic}, 2.24\cite{zhao2000electronic}, 2.59\cite{wenzien1995quasiparticle} &2.360\cite{levinshtein2001properties}\\
    2H  &   2.335   &   2.350\cite{huang2015thermoelectric}&   3.311   &  3.33\cite{brudnyi2012electronic}  &3.330\cite{patrick1966growth}  \\
    4H  &   2.244   &   2.238\cite{huang2015thermoelectric}&   3.205   &  3.26\cite{brudnyi2012electronic}, 3.11\cite{zhao2000electronic} &3.230\cite{levinshtein2001properties}  \\   
    6H  &   2.050   &   2.034\cite{alkhaldi2019crystal},2.031\cite{huang2015thermoelectric}&   3.127     &  3.05\cite{brudnyi2012electronic}  &3.000\cite{levinshtein2001properties}\\
    15R &   1.977   & 2.16\cite{zhang2014first}&   3.062       & - &2.986\cite{humphreys1981wavelength} \\ \hline
    \end{tabular}
    \label{tab:bg}
\end{table*}

We next show that our calculated electronic band gaps agree well both with both experiment and with previous calculations for all SiC polytypes. The calculated band gaps of the different SiC polytypes are listed in \ref{tab:bg}.
The PBE results exhibit the well-known underestimation of the band gap for all polytypes. However, including quasiparticle corrections with the $GW$ approximation, the calculated band gap is in much better agreement with experimental measurements, with the differences being within 0.15 eV. Our results exhibit a consistent trend as other theoretical and experimental studies: the indirect gap of the material decreases as the number of hexagonal layers increase. Meanwhile, the 3C structure shows a band gap about 1 eV narrower than all other polytypes, while the 15R structures show slightly smaller band gap compared to the 6H structure. It can be seen, however, that our calculated $GW$ band gap does not consistently overestimate or underestimate the experimental gap. The difference between our $GW$ result in comparison to other literature values may result from the plasmon-pole model, the different starting point due to the exchange-correlation functional, etc. The difference between the calculated value from the $GW$ approximation and the experimentally measured value is attributed in part to the lack of temperature-related renormalization ($\sim$ 0.03-0.05 eV\cite{choyke1969optical}) of the band gap and the zero-point motion. Nevertheless, the differences between our calculated $GW$ band gap and experimental measurements are in general within 0.15 eV, with the 3C structure being the largest but not exceeding 7\%. We note that fine tuning the electronic-structure methodology to obtain a better match of the gap to experiment is not the focus of this work, thus we restrict our band-structure calculation to the one-shot $GW$ approximation.

\begin{figure*}[!ht]
\centering
\includegraphics[width=\textwidth]{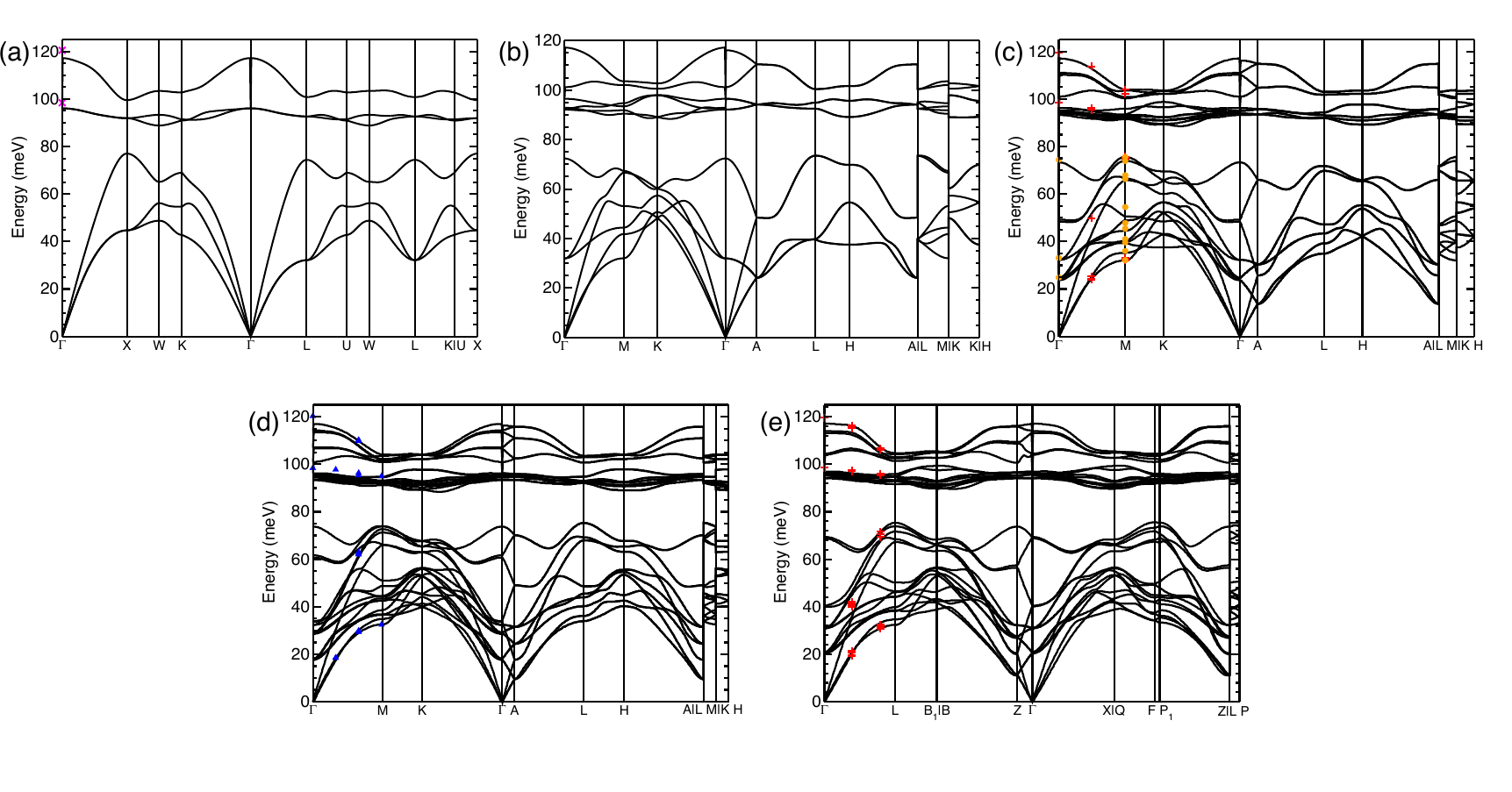}
\vspace{-1.5 cm}
\caption{\label{fig:phbands}Calculated phonon dispersions of the investigated SiC polytypes (a) 3C, (b) 2H, (c) 4H, (d) 6H, and (e) 15R. Our calculated phonon frequencies underestimate available experimental measurements by less than 3\%. Experimental data are plotted from: Ref. \cite{NIENHAUS1995L328} (3C, purple X symbols), Ref. \cite{feldman684H15r} (4H and 15R, red crosses), Ref. \cite{bai2002four} (4H, orange circles) and Ref.\cite{feldman68} (6H, blue upper triangles). Compared to previous theoretical studies\cite{petretto2018high}, we find very good agreement of the dispersion compared to experiment, with maximum differences of the phonon frequencies on the order of 2\%. 
}
\end{figure*}

We further show the calculated phonon band structures, and we show that the phonon frequencies agree well with both experimental measurements as well as theoretical studies. The calculated phonon band structures for all different SiC polytypes are shown in Figure \ref{fig:phbands}. The maximum phonon frequency at $\Gamma$ is very similar across all different polytypes, with a variation within the range from 117.0 meV to 117.2 meV. Compared to available experimental data (See marks for the data and references in Figure \ref{fig:phbands}), the differences are less than 3\%. We also compared the calculated phonon frequencies to first-principles results\cite{petretto2018high} using ABINIT\cite{GONZE2002478}, and the differences of the calculated phonon frequencies are no larger than 2\%. The calculated electron phonon matrix elements are interpolated onto a fine q and k-grid utilizing Wannier interpolation to obtain converged phonon-assisted optical spectra that we show later in section \ref{sec:indabs}.

\subsection{\label{sec:dirabs}Direct optical properties}

\begin{figure}[!ht]
    \centering
    \includegraphics[width=\columnwidth]{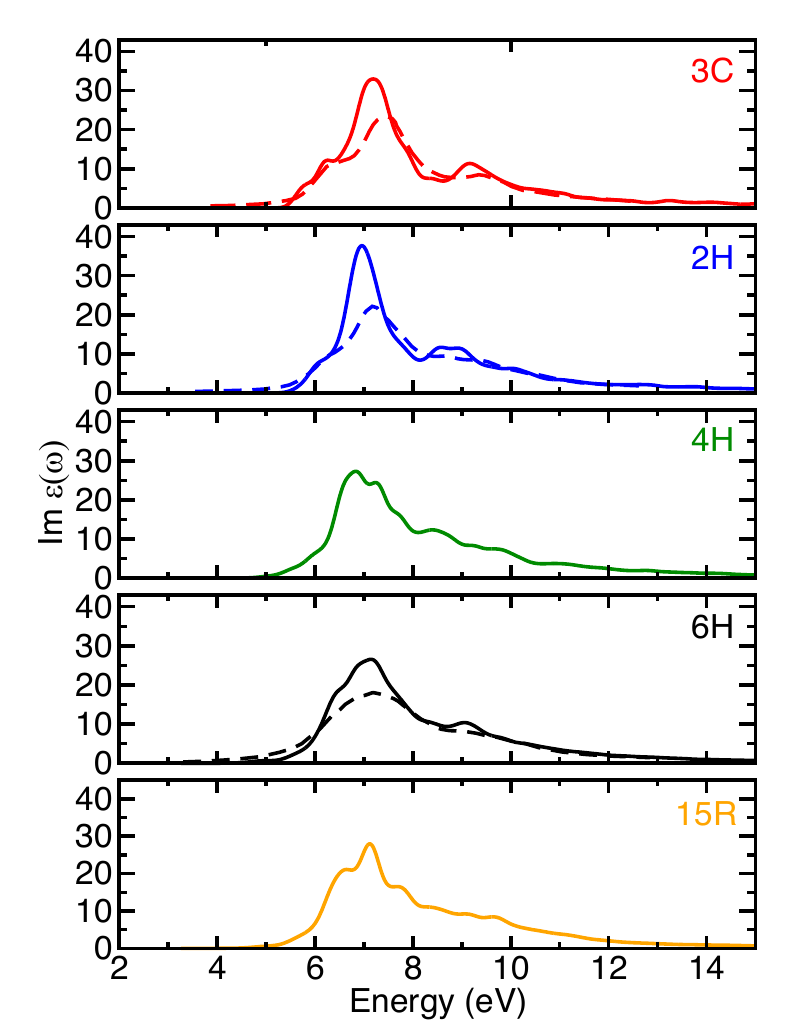}
    \caption{Solid: Calculated direct part of the optical absorption spectra (imaginary part of the complex dielectric function) for the ordinary light polarization for the five different SiC polytypes, including quasiparticle effects with the $GW$ approximation and electron-hole interactions via the BSE equation. Dashed: Theoretical BSE data from Ref.\cite{rohlfing2001} for 3C, 2H and 6H polytypes. We find very good agreement in terms of the peak positions and shapes, with the height of the peak potentially affected by the choice of the broadening parameters.
    }
    \label{fig:direct}
    \vspace{-0.5cm}
\end{figure}

We first report the calculated direct part of the absorption spectra and the dielectric constants, with excitonic effects included. The optical spectra shown in Figure \ref{fig:direct} (ordinary direction for the hexagonal structures) are calculated with quasiparticle energies from the $GW$ approximation and with electron-hole interaction from the BSE approach. Our calculated optical spectra agree well with literature\cite{rohlfing2001} for the 3C, 2H and 6H structure, especially considering the relative height and position of the main peak. Similar to Ref.\cite{rohlfing2001} we observe a significant shoulder close to the absorption onset ($\sim$6 eV) that is clearest for 3C SiC, while being less significant for 2H and 6H. It is worthwhile to note that our spectra are noticeably higher than in Ref.\cite{rohlfing2001}, which is due to the different choice of broadening parameters used to approximate the delta function (0.15 eV in this work versus 0.25 eV in the reference). We report the calculated dielectric constants in Table \ref{tab:eps}. Our calculated dielectric constants are higher than Ref.\cite{rohlfing2001} and this can be affected by many factors, including possible different BSE energy cutoffs and different quasiparticle band gaps, etc. Overall, the differences between our calculated dielectric constant and experimental values are within 5\%. Since in the indirect part of the spectra, the imaginary part of the dielectric functions are orders of magnitudes smaller than the direct part, it is reasonable to assume that the refractive index is approximately proportional to the square root of the real part of the dielectric function. As a result, in this case the difference in the calculated direct spectra from theory to experiment does not affect the calculated indirect part of the spectra by more than 5\%. Later in this section, we show that this is a minor effect compared to other factors, for example, the finite-temperature band-gap renormalization, or the mere difference of the band gap from $GW$ approximation to experiment.

\begin{table}
    \centering
    \caption{Dielectric constant for the ordinary polarization ($\varepsilon^{\perp}_{\infty}$) calculated with the BSE approach for the five SiC polytypes. Our calculated dielectric constants overestimate experimental values by less than 5\%, resulting in less than 3\% overestimation for the refractive index. 
    }
    \setlength{\tabcolsep}{4.5pt}
    \begin{tabular}{cccc}
    \hline Polytype & \makecell{This work,\\BSE}&\makecell{Theory, \\BSE (Ref.\cite{rohlfing2001})}&\makecell{Experiment \\(Ref.\cite{harris1995properties})} \\ \hline
        3C & 6.93 & 6.36 &6.52 \\
        2H & 6.89 & 6.27 &6.51\\ 
        4H & 6.84 & - & -\\
        6H & 6.79 & 6.31 &6.52 \\
        15R & 6.73 & - & -\\ \hline
    \end{tabular}
    \label{tab:eps}
\end{table}

\subsection{\label{sec:indabs}Phonon-assisted optical properties from second-order perturbation theory}

\begin{figure}[!ht]
    \centering
    \includegraphics[width=\columnwidth]{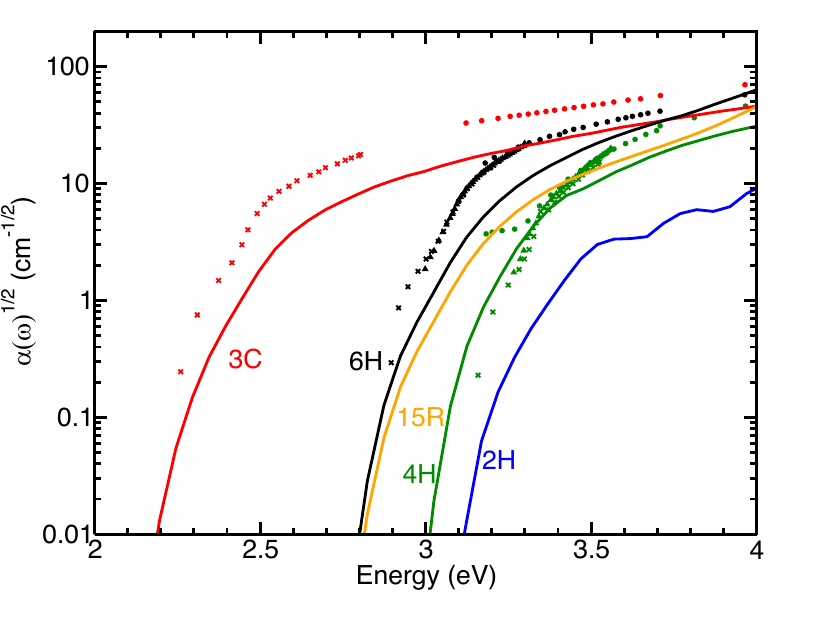}
    \vspace{-1cm}
    \caption{Calculated optical absorption coefficient for the ordinary direction (E$\perp c$) at 300 K for the five investigated SiC polytypes in the phonon-assisted spectral region between the indirect and direct band gaps. The experimental results from the literature are taken from Ref.\cite{vsvcajev2011diffraction} (3C, 4H and 6H, cross mark), Ref.\cite{sridhara1999penetration} (4H and 6H, dot marks) and Ref.\cite{watanabe2014temperature} (4H and 6H, upper triangle mark). A rigid shift has been applied to each theoretical curve to correct the difference between the GW-calculated gaps and the experimental values at 300 K listed in Table \ref{tab:bg} for each polytype. The calculated absorption spectra are in overall good agreement with experiment.
    }
    \label{fig:indabs}
\end{figure}

Following the calculation of the direct spectra, we determine the phonon-assisted indirect optical spectra of the SiC polytypes, evaluated with the second-order time-dependent perturbation theory. We find that our results agree well with experiments for the most common polytypes (3C, 4H and 6H), and we can predict the indirect spectra for 2H and 15R polytypes. Figure \ref{fig:indabs} shows the calculated absorption coefficient in the region of indirect absorption for all polytypes. Additional rigid shifts $\Delta$ are applied to the quasiparticle energies from the $GW$ approximation to match the indirect gap from experiment: $\Delta=-0.154$ eV for 3C, $\Delta=0.019$ eV for 2H, $\Delta=0.025$ eV for 4H, $\Delta=-0.126$ eV for 6H and $\Delta=-0.076$ eV for 15R. For the indirect part of the spectra, it is reasonable to assume that the correction on the band gap can be well represented with a rigid shift of the absorption coefficient (i.e., no scaling of the spectra is needed).\cite{zacharias2016one} We compared the calculated absorption coefficient in the indirect region of the 3C, 4H and 6H polytypes to experimental measurements, and we find a good agreement with the experimental results. For the 15R structure, we observe that the absorption coefficient in the indirect region is similar to the 6H structure. Such similarity is expected due to the similarities in the electronic band structure and phonon band structures. Interestingly, for the 2H structure, we observe two bump-like features around photon energies of 3.45 eV and 3.75 eV. We note that these two energies correspond to the energy difference between the valence band maximum at $\Gamma$ and the conduction band minimum at $K$, as well as the second minimum at $M$, respectively. This is a unique feature that we only observe in the 2H structure, as for all other hexagonal structures, the CBM is located at $M$, rather than $K$.

\begin{figure}[!ht]
\vspace{0.3cm}
\includegraphics[width=\columnwidth]{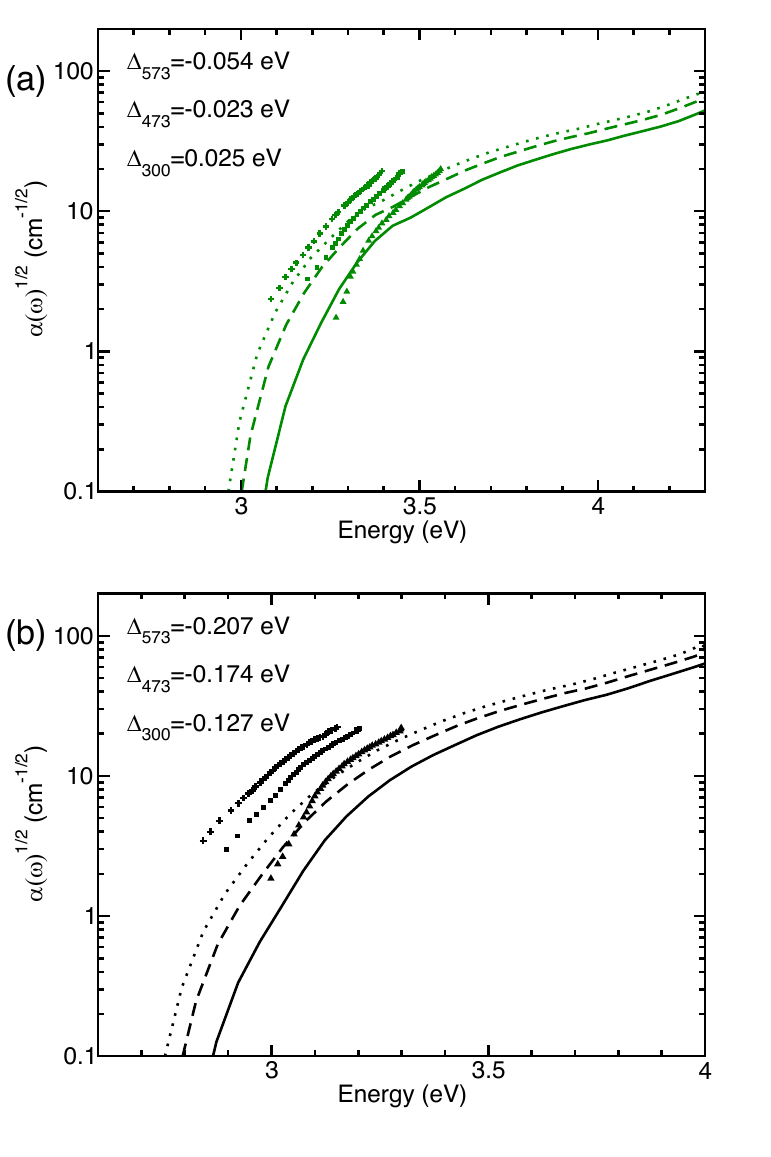}
\vspace{-1cm}
\caption{\label{fig:temp_indabs} Calculated absorption coefficient for (a) 4H SiC and (b) 6H SiC as a function of photon energy in the phonon-assisted region (lines) for a temperature of 300 K (solid), 473 K (dashed), and 573 K (dotted), and compared to experimental measurements (data points) from Ref.\cite{watanabe2014temperature} at 300 K (upper triangle), 473 K (square) and 573 K (plus). The theoretical curves have been rigidly shifted to match the temperature dependence of the experimental band gap according to Ref.\cite{levinshtein2001properties}. The change in the spectra with respect to temperature change is well-predicted with the changes in band gaps included.
}\vspace{-0.2cm}
\end{figure}

Stepping forward, our first-principles tools allow us to investigate the temperature dependency of the indirect optical properties to examine the factors that affect the spectra as temperature changes. As a thermally stable material, many of the novel optoelectronic applications of SiC involve extreme conditions.\cite{lohrmann2017review,kazakova2018analysis,ebrahimi2020innovative} Therefore, being able to predict temperature-dependent optical properties from a first-principles perspective can provide crucial information for novel applications. We calculated the temperature-dependent indirect optical spectra for the 4H and 6H polytypes at three different temperatures (300 K, 473 K and 573 K). The calculated temperature-dependent absorption coefficients in the indirect region are shown in Figure \ref{fig:temp_indabs}. Additional rigid shifts are applied to match the calculated electronic band gap with experimentally determined values.\cite{levinshtein2001properties} It can be seen from the figure that our calculation reproduces the relative change in the indirect optical absorption due to temperature changes very well after taking the band gap renormalization into account (See appendix section \ref{secapp:temp} for spectra without additional rigid shift). We conclude that the effects of temperature on both the phonon occupation numbers as well as on the renormalization of the band gap itself are important to obtain reliable prediction of the indirect absorption spectra. However, for the 6H polytype, we find that the simple rigid shift of the gap does not accurately provide the correct positions of the shoulder of the absorption curve. In the next section, we examine an alternative route to calculate phonon-assisted absorption that avoids the rigid-shift approach and considers implicitly the temperature renormalization of the band structure.  

\subsection{Phonon-assisted optical properties from the special displacement method (SDM)}

\begin{figure}
    \centering
    \includegraphics[width=\columnwidth]{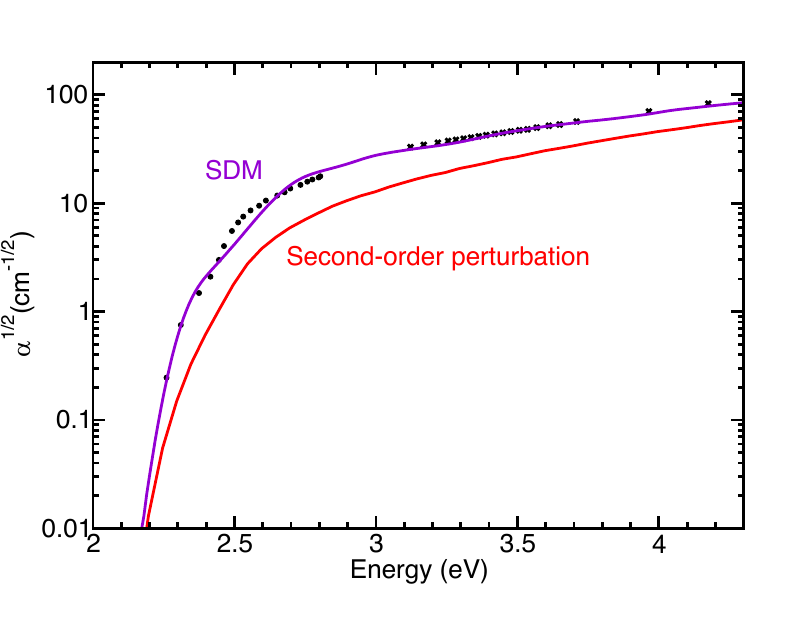}
    \vspace{-1cm}
    \caption{Calculated absorption coefficient as a function of energy for the 3C polytype with $6\times6\times6$ supercell with the special displacement (SDM) approach\cite{zacharias2020theory} using a $4\times4\times4$ $\mathbf{k}$-points grid (violet) and the perturbation approach (red solid) ($32\times32\times32$ $\mathbf{k}$/$\mathbf{q}$-points grid). Experimental data from Ref.\cite{vsvcajev2011diffraction} (black dots) and Ref.\cite{sridhara1999penetration} (black cross) are shown for comparison. The same rigid shift is applied as in Figure \ref{fig:indabs} for the perturbation approach, while no rigid shift is applied for the SDM approach. The renormalization of the band gap due to temperature is physically included with the SDM approach.
    }
    \label{fig:hbb_wl}\vspace{-0.2cm}
\end{figure}

Next, we show that the effect of temperature-related renormalizations of the band gap on the optical absorption spectra of SiC can be overcome by applying the special displacement method. Although the SDM approach only requires a single snapshot of the atomic displacement configuration,\cite{zacharias2016one,zacharias2015stochastic, zacharias2020theory} the need to perform calculations on relatively large supercells (approaching thousands of atoms) results in a drastic increase in the computational cost. As a result, we take quasiparticle corrections into account using a rigid increase of the gap (1.141 eV) to account for the difference between the PBE band gap and the GW band gap from unit cell calculations. In Figure \ref{fig:hbb_wl}, we show the best converged results using the SDM approach compared to the perturbation approach for 3C SiC. The same rigid shift is applied for the perturbation approach as indicated in the previous sections. In the SDM approach, no artificial shifts are applied except for the indirect-band-gap  correction from the PBE to the GW value. At energies beyond 3 eV, the SDM approach shows better agreement with experiment quantitatively compared to experimental data. This is because, by taking the renormalization of the band gap with respect to temperature into account, the SDM approach is more physical in terms of predicting the correct position of the absorption peaks. However, below a photon energy of 3 eV, the perturbation approach predicts the shape of the curve better. There are a few possible factors: First, it is straightforward to use the perturbation approach utilizing Wannier interpolation to obtain matrix elements on the fine grid, which is important in order to obtain converged spectra near the absorption onset that require a fine sampling of the Brillouin zone. For the SDM approach, however, convergence with respect to supercell size is required. In our work, we find that under-converged supercell can incur significant error for the absorption onset, or incorrect absorption shoulders, and a $6\times6\times6$ supercell is needed for converged phonon-assisted spectra (See appendix section \ref{secapp:convzg}). Second, the nature of the approach corresponds to neglecting the explicit phonon frequencies in the delta-function of calculating $\varepsilon_2(\omega)$ in Eq.\eqref{eqn:eps2}, thus, the absorption onset can be affected.\cite{zacharias2016one,zacharias2015stochastic, zacharias2020theory}

Overall, we show that although both the SDM approach and the perturbation approach provide satisfying results for predicting the phonon-assisted optical absorption properties for 3C SiC, the SDM approach clearly illustrates the importance of including the correct temperature-dependent renormalization of the band gap to obtain higher quantitative accuracy. The special displacement approach also shows the potential of further investigations, e.g., excitonic effects on the indirect part of the optical spectra by performing BSE calculations on properly converged supercells. However, fully converged BSE calculations on such large supercells is computationally challenging and is beyond the scope of this paper. Meanwhile, it is also clear that, to resolve the finer structures of the optical absorption around the absorption edge, it is important to fully consider the effects of phonons including the phonon energies.

\section{\label{sec:conclusion}Conclusion}
In this work, we investigate the phonon-assisted optical properties of five different SiC polytypes with first-principles calculations. We first calculate the structural properties of all polytypes with density functional theory, and electronic properties with the GW approximation, and we show that the calculated results agree well with both other theoretical studies and experimental measurements. We then utilize Wannier interpolation to obtain the electron-phonon coupling in dense electronic and phonon grids, and we use second-order time-dependent perturbation theory to obtain phonon-assisted optical spectra in the indirect gap region of all polytypes. We show that the simulated absorption coefficient in the indirect region agrees well with experimental measurements for the most common 3C, 4H and 6H polytypes, and provides prediction for 15R and 2H polytypes. Further analysis on the temperature-dependent indirect optical properties shows that our theoretical simulation predicts the trend of the variation in optical properties with respect to temperature well, and we show that our simulations can be particularly useful in understanding the practical application of the material at various finite temperatures. However, our results shows that the renormalization of the band gaps due to temperature is important in predicting the temperature-dependent spectra. Therefore, we further compare the conventional approach utilizing second-order perturbation theory to the method employing special displacement of specific atoms in supercells, and we show that improvement of the predictions of the temperature dependency of the indirect optical spectra can be obtained by take the renormalization of the electronic energies implicitly due to temperature change into account. Our work shows that these methods enable quantitively predictions of temperature-dependent phonon-assisted absorption spectra in indirect-gap semiconductor materials.  

\section*{Acknowledgement}

We thank Dr. Marios Zacharias for the fruitful discussion about the SDM approach. The work is supported as part of the Computational Materials Sciences Program funded by the U.S. Department of Energy, Office of Science, Basic Energy Sciences, under Award No. DE-SC0020129. Computational resources were provided by the National Energy Research Scientific Computing Center, which is supported by the Office of Science of the U.S. Department of Energy under Contract No. DE-AC02-05CH11231.

\section{Appendix}
\subsection{Refractive index}\label{secapp:nr}
\vspace{-0.5cm}
\begin{figure}[!ht]
    \centering
    \includegraphics[width=\columnwidth]{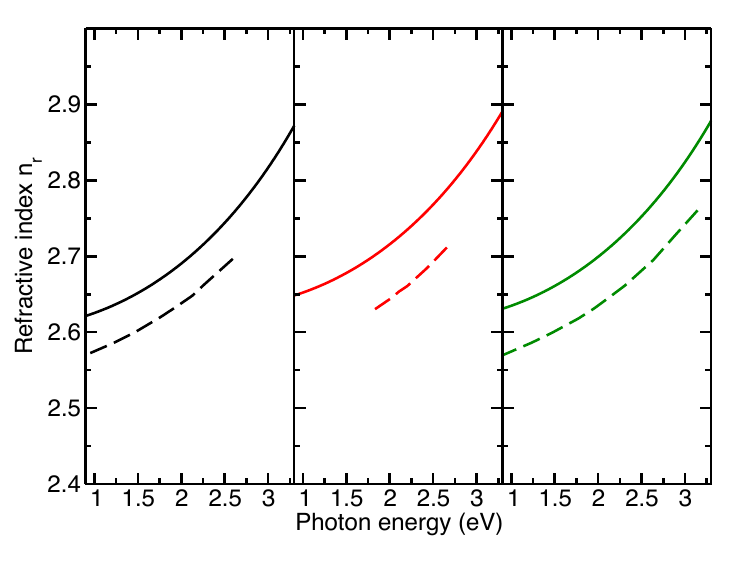}\vspace{-0.6cm}
    \caption{Calculated refractive index in the region close to the indirect band gap for 4H SiC (black solid), 3C SiC (red solid) and 6H SiC (green solid), as well as comparison to experimental curve (dashed) from Ref.\cite{wang20134h} (4H and 6H) and Ref.\cite{Shaffer:71} (3C).  }
    \label{fig:ref_index}
\end{figure}

In this section we show the calculated refractive index compared to experimental measurements. Figure \ref{fig:ref_index} shows that our calculated refractive index of 3C, 4H and 6H polytypes agrees very well with experimental measurements, with the difference being an overestimation by less than 3\%. These differences in the refractive index from our calculation compared to experiments are a negligible source of error even compared to the mere difference of the band gaps themselves.

\subsection{Temperature dependent indirect spectra}\label{secapp:temp}
\vspace{-0.8cm}
\begin{figure}[!ht]
\includegraphics[width=\columnwidth]{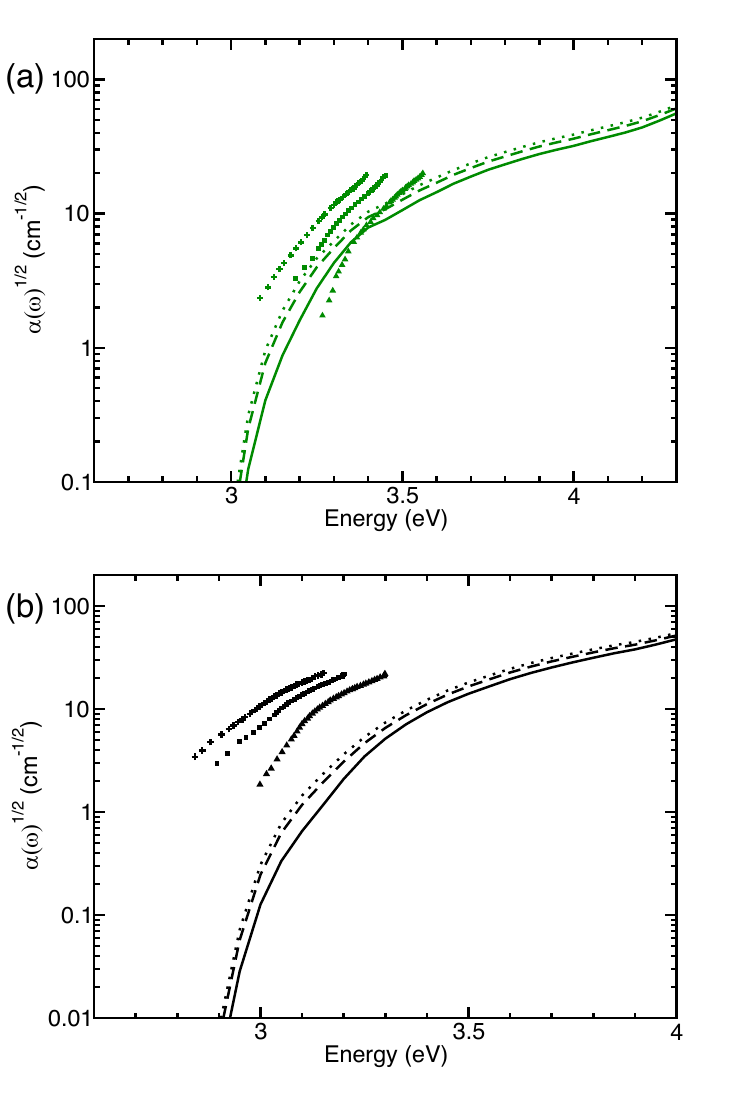}\vspace{-0.6cm}
\caption{\label{fig:s:temp_indabs} Calculated absorption coefficient for 4H SiC (a) and 6H SiC (b) as a function of photon energy in the indirect region (lines) for a temperature of 300 K (solid), 473 K (dashed), and 573 K (dotted), and compared to experimental measurements (data points) from Ref.\cite{watanabe2014temperature} at 300 K (upper triangle), 473 K (square) and 573 K (plus). The difference with Figure \ref{fig:temp_indabs} is the omission of the correction to the band-gap value to match experiment.
}
\end{figure}

In this section, we demonstrate the temperature-dependent indirect spectra without including the additional rigid shifts to account for the temperature-dependent band-gap renormalization. In Figure \ref{fig:s:temp_indabs}, we show the calculated optical spectra without considering the difference between calculated band gap and experimentally measured band gap. It can be seen clearly that first, the effects of temperature on the indirect optical spectra itself is still clear, and this is mainly due to the change in Bose-Einstein occupation factor of the phonons. However, without considering the difference in band gap from theory and from experiment, especially in the 
6H case, the onset of absorption is severely overestimated, and more importantly in this context, the differences between the curves at different temperatures is underestimated. This clearly shows the importance of considering both the change in occupations due to temperature and the changes in band gap itself due to temperature change, as the change in spectra is clearly a combined effect of both.
\subsection{Convergence of the supercell approach}\label{secapp:convzg}
\vspace{-0.8cm}
\begin{figure}[!ht]
    \centering
    \includegraphics[width=\columnwidth]{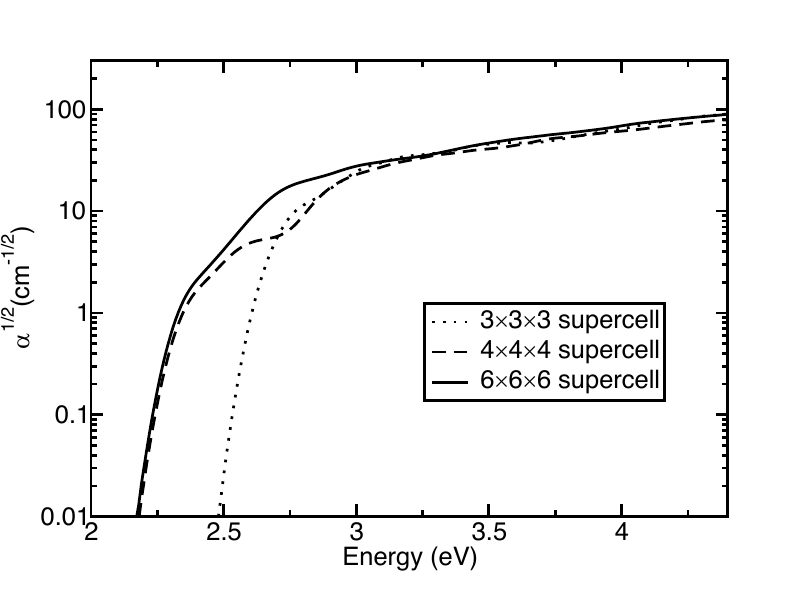}\vspace{-0.6cm}
    \caption{Calculated absorption coefficient using the special displacement approach\cite{zacharias2020theory} with three different supercell sizes: $3\times3\times3$ (dotted), $4\times4\times4$ (dashed) and $6\times6\times6$ (solid). 
    The $\mathbf{k}$-points sampling are $8\times8\times8$, $6\times6\times6$ and $4\times4\times4$, respectively. All k-point grids are randomly shifted for better convergence. Convergence beyond $6\times6\times6$ cell require supercells with more than 1000 atoms and is not tested due to the large computational cost.
    }
    \label{fig:s:conv_ZG}
\end{figure}
To test the convergence of the supercell approach versus the supercell size, we calculated the indirect optical spectra with $3\times3\times3$, $4\times4\times4$ and $6\times6\times6$ supercells. In this section we report the convergence behavior of the calculated spectra near the absorption onset in Figure \ref{fig:s:conv_ZG}. It can be seen from the figure that the two smaller supercells are not sufficient for converged optical spectra in the indirect region. The best converged results of the $6\times6\times6$ supercell is reported in the main text.

\bibliography{apssamp}

\end{document}